\documentclass[twocolumn,british,english]{aastex6}

\usepackage{units}
\usepackage{amsmath}
\usepackage{wasysym}

\begin{document}

\title{Performance of an Algorithm for Estimation of Flux, \\ 
Background and Location on One-Dimensional Signals }
\shorttitle{Performance of an algorithm for estimation of 
flux, background and location }

\author{M. Gai \altaffilmark{1}, D. Busonero \altaffilmark{1} \and 
R. Cancelliere \altaffilmark{2}} 
\affil{
\altaffilmark{1}Istituto Nazionale di Astrofisica, Osservatorio 
Astrofisico di Torino, V. Osservatorio, 20 - 10025 Pino Torinese, Italy 
\\ 
\altaffilmark{2}University of Turin, Department of Computer Sciences, 
V. Pessinetto 12 - 10149 Torino, Italy
}
\shortauthors{Gai, Busonero and Cancelliere}


\begin{abstract}
Optimal estimation of signal amplitude, background
level, and photocentre location is crucial to the combined extraction of
astrometric and photometric information from focal plane images, and
in particular from the one-dimensional measurements performed by {\it Gaia} 
on intermediate to faint magnitude stars. 
Our goal is to define a convenient maximum likelihood framework, suited 
to efficient iterative implementation and to assessment of noise level, 
bias, and correlation among variables. 
The analytical model is investigated numerically
and verified by simulation over a range of magnitude and background
values. 
The estimates are unbiased, with a well-understood correlation between 
amplitude and background, and with a much lower correlation of either of them 
with location, further alleviated in case of signal symmetry. 
Two versions of the algorithm are implemented and tested against each 
other, respectively, for independent and combined parameter estimation. 
Both are effective and provide consistent results, but the latter is 
more efficient because it takes into account the flux-background 
estimate correlation. 
\end{abstract}

\maketitle

\section{Introduction }
\label{sec:intro}
Astrometric measurements are often concerned with the limiting precision
achievable in the estimate of relative position among celestial sources,
imaged by some instrument, and with the practical definition of convenient
location algorithms \citep{Lindegren1978,GaiPASP1998}. Such mathematical
frameworks often identify convenient quantities, in particular an
assessment of the limiting achievable precision, e.g. in terms of
the Cram{\'{e}}r-Rao lower bound on location error \citep{Mendez2013},
or of systematic errors associated to the mismatch between the measurement
model and actual data distribution \citep{GaiPASP2013Fit}. 
The connection between maximum likelihood (ML), related to Cram{\'{e}}r-Rao, 
and least square approach is also investigated in the literature \citep{Mendez2015}. 
The models have been tested in real observations, even from the ground 
\citep{Cameron2009}, where the atmospheric turbulence is usually the limiting 
factor, and in the lab \citep{GaiAA2001}, sometimes to very high precision 
\citep{ZhaiShao2011}. 

Often the problem formulation is simplified with the adoption of a
one-dimensional (1D) signal model, which is more directly applicable,
e.g., to the {\it Gaia} measurements over its intermediate to faint magnitude
sample, which is briefly reviewed in Sec.\ref{sec:GaiaMission}, but can nonetheless
be considered sufficiently general, in particular by dealing separately
with image location on each coordinate. Besides, the problem is sometimes
more conveniently formulated in terms of estimating not only the image
location, but also the photometric level at the same time, i.e. working
in a bidimensional space with respect to the unknowns \citep{Mendez2014}.
The rationale is that, even in case of known objects, the actual photon
count accumulated in a specific exposure is dependent on the current
measurement conditions (atmospheric transmission, effective exposure
time, and instrument parameters), which may not be stable or known with
accuracy adequate to the precision goals. 

In astronomical practice, measurement precision is affected also by
background, which is in many cases estimated over image regions close
to the object of interest and considered free from residual target
flux or spoilers. Background estimation can be performed in many different
ways, depending on a number of assumptions related to the actual
observing conditions and goals, and requires careful evaluation. 
It is therefore a critical part of several data reduction and analysis
packages, e.g. the public domain AstroImageJ \citep{AstroImageJ2016}. 

The subject of this paper is an algorithm for the estimation of signal
Amplitude, Background, and photoCentre location (ABC) on 1D 
data corresponding to intermediate magnitude {\it Gaia} observations. 
We deal, therefore, with a three-dimensional (3D) problem, in the sense of \citet{Mendez2014},
formulated in Sec.~\ref{sec:MaxLikApproach} in maximum likelihood terms. 
The expressions are expanded in a form suited to an iterative solution, 
assuming the variables are either independent or a combined set. 
These two approaches are materialized in different algorithms, 
and their noise and correlation properties are derived, evidencing 
relevant consequences for symmetric signals. 

Because the mathematical expressions involved cannot be easily solved
except in the case of extremely simple signal models, the framework
is verified by numerical simulation as described in Sec.~\ref{sec:Simulation},
also taking into account some of the aspects more relevant toward
implementation in a data reduction system \citep{Morbidelli2012}. 

The two algorithms have comparable performance with respect to noise 
and systematic error, but the correlation between flux and background, 
discussed in Sec.~\ref{sec:Discussion}, makes the independent estimate 
significantly less efficient.

\subsection{{\it Gaia} Observations }
\label{sec:GaiaMission}
The {\it Gaia} mission 
\citep{GaiaDR1,Perryman2005,deBruijne2012} 
is aimed at global astrometry at the level of a few tens of micro-arcsec 
(hereafter, $\mu as$), producing an all-sky catalogue 
of position, proper motion, and parallax, complete to the limiting
magnitude $V \simeq 20\,mag$. 
The {\it Gaia} concept (http://sci.esa.int/gaia/)
relies on self-consistency of the astrometric information of celestial
objects throughout operation, factoring out the instrument parameters
and their evolution by calibration of the overall data set. 
The {\it Hipparcos} experience suggests that the approach is viable with 
respect to the detection 
and modeling of the instrumental parameter secular evolution and long
term variations over time scales longer than a few revolution periods,
i.e. above about one day. 

The detector of {\it Gaia} \citep{Short2005,Kohley2012} is a large CCD mosaic
operated in Time Delay Integration (TDI) mode, split in different
regions for operation purposes: the detection of the incoming stars by
the initial two strips of CCDs, the Sky Mapper; wideband imaging on
the Astrometric Field; and low dispersion imaging on the Blue/Red Photometer.
Additionally, a fraction of the field feeds the Radial Velocity Spectrometer, 
operating over the magnitude range $G \apprle 17 \, mag$, 
which also provides astrophysical characterization of the bright stellar
sample ($G \apprle 12.5 \, mag$). 
The Basic Angle Monitoring device provides auxiliary metrology information, 
i.e. a real-time estimate of the Basic Angle 
value\citep{Gielesen2012,GaiPASP2013BAM}. 

The regions of interest over the astrometric focal plane are defined by 
onboard logic on the detection results in the Sky Mapper: for each 
detected star, the placement of ensuing observing windows on each CCD is 
computed. 
\\ 
The elementary exposure has a fixed duration ($\sim4\,s$) and provides 
for most intermediate magnitude stars ($13\leq G\leq16\,mag$) a 1D 
data set of 18 samples, binned in the low resolution, across scan 
direction. Fainter stars are binned and read on 12 samples. 
Brighter stars, in the magnitude range $11.8\leq G\leq13\,mag$, 
are read and downloaded as bidimensional elementary images. 
The brightest stars, $G\leq11.8\,mag$ are read on shorter exposure times, 
reducing the effective integration by activation of on-chip gates.

\subsection{Estimate of Flux, Background and Photocentre }
The elementary 1D signal depends on the instrument response (including
optics, TDI, attitude, and CCD) and on the individual source magnitude
and spectrum. The apparent amplitude of a star is thus slightly different
on each exposure during the transit, due to different electro-optical
response of subsequent CCDs. Keeping track of the star flux level
over the AF is therefore a valuable contribution to instrument monitoring,
as many stars are continuously crossing the field at different
heights, each sampling a whole CCD strip. The effective broadband
magnitude of each star can be defined, e.g., by its average signal level
over the whole transit. 
Source variability, if any, is most likely identified between observations 
at different epoch. 

Moreover, due to field superposition from the two {\it Gaia} telescopes,
the background from both sky areas is superposed to the image of each
star, so that every star is measured over a different background at
each epoch. The estimation of background is therefore required at
the individual exposure level, since it is not known in advance or
easily modeled. 

Finally, the most obvious parameter desired for each elementary exposure
on the astrometric field is the star location, i.e the estimate of
its photocentre. 

The next section is devoted to setting up the mathematical framework
describing the maximum likelihood search for photometric level, background
and image location, in a formulation convenient for iterative numerical
implementation.

\section{Method: Maximum Likelihood }
\label{sec:MaxLikApproach}The signal is one-dimensional, centred
on the ``true'' photoCentre position ($C$) and represented in pixel
positions $x_{k}$, $k=1,\ldots,K$ by the model 
\begin{equation}
U_k =U\left(x_{k};A;B;C\right)=A\cdot T\left(x_{k}-C\right)+B\,,\label{eq:SignalModel}
\end{equation}
where $T$ is the flux independent Template (normalised with unit
integral), $A$ is the Amplitude of the total photometric flux, and
$B$ is the Background, assumed to be uniform. 
The signal template is derived from sufficiently large, 
homogeneous data sets (e.g. from the same CCD and spectral class) in 
calibration \citep{GaiPASP2013Fit,Busonero2014}, to ensure that a good 
Signal to Noise Ratio (SNR) is achieved. 

The elementary exposure generates the actual data samples $S_{k}$,
affected by photon noise from the signal and background, readout noise,
and possibly other contributions. 
The \textit{discrepancy} between the detected signal and 
its model is then 
\begin{equation}
d{}_{k} = S_k - U_k = S_{k}-A\cdot T\left(x_{k}-C\right)-B\,,
\label{eq:SigDiscr}
\end{equation}
assumed \textit{unbiased} and \textit{uncorrelated},
i.e., respectively, 
\begin{equation}
\left\langle d_{n}\right\rangle =0\,;\,\,\left\langle d_{m}d_{n}\right\rangle 
= \delta_{mn}\sigma_{n}^{2}\,,\label{eq:SigStats}
\end{equation}
where $\sigma_{k}^{2}$ is the signal variance, assumed to be known. 

The estimate of centroid, flux and background can be implemented in
a maximum likelihood framework by the requirement of minimising an
error function (corresponding to the $\chi^{2}$ in other applications)
chosen as the weighted square \textit{discrepancy} $D$: 
\begin{equation}
D=\sum_{k=1}^{K}\frac{\left[S_{k}-A\cdot T\left(x_{k}-C\right)-B\right]^{2}}
{\sigma_{k}^{2}}\,.\label{eq:ChiSquare}
\end{equation}
Hereafter, for simplicity, the index $k$ will be dropped; 
the summations cover the whole pixel range. 

The solution minimises the sum of squared differences between samples
and corresponding model values. 
It may be noted that the signal model
formulation in Eq.~\ref{eq:SignalModel} is such that the solution
cannot be strictly derived in terms of \textit{linear least squares},
because, e.g., amplitude and photo-centre are intrinsically coupled. 
As often happens, it is possible to envisage an \textit{iterative solution},
based on an acceptable \textit{initial guess} of the parameters, and
successive approximations providing progressively better estimates
at each iteration. 
Given the simple signal profile, i.e. a bell-shaped curve affected 
by moderate noise for a reasonable SNR, the 
problem is still expected to have a single optimum solution. 
Moreover, the least square solution often provides a good approximation 
to the maximum likelihood solution and to the Cram{\'{e}}r-Rao 
limit \citep{Mendez2015}. 

The stationary point of the error functional satisfies the system
of equations: 
\[
\frac{\partial D}{\partial A}=0\,;\,\frac{\partial D}{\partial B}=0\,;\,
\frac{\partial D}{\partial C}=0\,,
\]
i.e. 
\begin{align}
\sum\frac{\left[S_{k}-A\cdot T\left(x_{k}-C\right)-B\right]\cdot 
T\left(x_{k}-C\right)}{\sigma_{k}^{2}}= & 0\nonumber \\
\sum\frac{\left[S_{k}-A\cdot T\left(x_{k}-C\right)-B\right]}{\sigma_{k}^{2}} = 
& 0\label{eq:derivChiSq}\\
\sum\frac{\left[S_{k}-A\cdot T\left(x_{k}-C\right)-B\right]\cdot 
T\,'\left(x_{k}-C\right)}{\sigma_{k}^{2}}= & 0\nonumber 
\end{align}
which can be solved with ordinary matrix methods after explicitation
of the variables. 

The three variables can be expanded at first order with respect to
the current (input) estimate, at iteration $\left(i\right)$, of amplitude,
background, and centre location $A_{E},\,B_{E},\,C_{E}$, to derive
the corrections $\delta A,\,\delta B,\,\delta C$ required to achieve
the improved (output) estimates $A_{N},\,B_{N},\,C_{N}$, i.e. 
\begin{equation}
A_{N} = A_{E}+\delta A\,;\, 
B_{N} = B_{E}+\delta B\,;\, 
C_{N} = C_{E}+\delta C\,.
\label{eq:Corrections} 
\end{equation}
The improved estimates, output of iteration $\left(i\right)$, become 
the current estimates fed as input to iteration $\left(i+1\right)$:
\[
A_{E}^{\left(i+1\right)}=A_{N}^{\left(i\right)},\,B_{E}^{\left(i+1\right)} = 
B_{N}^{\left(i\right)},\,C_{E}^{\left(i+1\right)}=C_{N}^{\left(i\right)}\,.
\]
In particular, the normalised template $T$ is expanded as 
\[
T\left(x_{k}-C_{N}\right)=T\left(x_{k}-C_{E}\right)-\delta C\cdot T\,' 
\left(x_{k}-C_{E}\right)\,.
\]
The updated version of the full signal model becomes 
\begin{eqnarray*}
U\left(x_{k};[N]\right) & = & U\left(x_{k};[E]\right)+\delta A\cdot 
T\left(x_{k}-C_{E}\right)\\
 &  & -A_{E}\delta C\cdot T\,'\left(x_{k}-C_{E}\right)+\delta B
\end{eqnarray*}

At each iteration, the problem can be tackled either by computing
the next estimate independently for each parameter, or collectively
for the set of three corrections at the same time; hereafter, the
two approaches are labeled Independent Estimate (IE)
and Combined Estimate (CE), respectively. 
The former approach is expected to provide a simpler formulation and 
easier connection with known results; besides, the latter approach is 
expected to be more sound, at the cost of more cumbersome expressions. 
Both are evaluated and discussed below, adopting for the sake of 
simplicity the notation: 
\begin{equation}
T\left(x_{k}-C\right) = T_k \, , \, T'\left(x_{k}-C\right) = T'_k \, . 
\end{equation}
The computational cost is evaluated in simulation. 

We take advantage of the maximum likelihood framework also to investigate
on the residual errors of the estimates, in particular with respect
to bias, variance, and correlation. In general, some correlation may
be expected due to derivation of all parameters from the same data,
and it can be justified intuitively by the characteristics of the
signal model in Eq.~\ref{eq:SignalModel}: e.g. given an underestimate
of the signal amplitude $A$, whichever the centre position $C$,
the overall photon count can be retrieved only by an overestimate
of the background $B$, i.e. the two parameters are anti-correlated.

\subsection{Independent Estimate (IE) }
\label{sub:IAS} Eqs.~\ref{eq:derivChiSq} are expanded
separately, respectively with regard to $\delta A,\delta B,\delta C$, i.e. to
the single variable of interest involved in the specific derivative
of the error function. Thus, we get the simple expressions 

\begin{eqnarray}
\sum\frac{\left[d_{k}-\delta A\cdot T_{k}\left(C_{E}\right)\right]\cdot 
T_{k}\left(C_{E}\right)}{\sigma_{k}^{2}} & = & 0\nonumber \\
\sum\frac{d_{k}-\delta B}{\sigma_{k}^{2}} & = & 0\label{eq:IndividExp1}\\
\sum\frac{\left[d_{k}+A_{E}\cdot\delta C \cdot T_{k}'\left(C_{E}\right)\right]\cdot 
T_{k}'\left(C_{E}\right)}{\sigma_{k}^{2}} & = & 0\nonumber 
\end{eqnarray}
where $d_{k}=d_{k}\left(x_{k};A_{E};B_{E};C_{E}\right)$; the equations
are solved independently for each variable, providing the corrections 

\begin{equation}
\delta A=\frac{\sum\nicefrac{d_{k}T_{k}}{\sigma_{k}^{2}}}
{\sum\nicefrac{T_{k}^{\,2}}{\sigma_{k}^{2}}}\,;\label{eq:IndividSol1}
\end{equation}

\begin{equation}
\delta B=\frac{{\displaystyle \sum\nicefrac{d_{k}}{\sigma_{k}^{2}}}}{{\displaystyle 
\sum\nicefrac{1}{\sigma_{k}^{2}}}}\,;\label{eq:IndividSol2}
\end{equation}

\begin{equation}
\delta C=-\frac{1}{A}\,\frac{{\displaystyle \sum\nicefrac{d_{k}T_{k}'}{\sigma_{k}^{2}}}}
{{\displaystyle \sum\nicefrac{T_{k}'^{2}}{\sigma_{k}^{2}}}}\,.\label{eq:IndividSol3}
\end{equation}

Some properties of the solution can be derived from the signal properties
in Eq.~\ref{eq:SigStats}. First, we remark that the Eq.~\ref{eq:Corrections}
can be used also to describe the discrepancy between the ``true''
value of each parameter and its current estimate at a given iteration,
including the last estimate; formally, the former takes the place
of the next estimates $A_{N},B_{N},C_{N}$. Then, the expressions
for the solutions in Eqs.~\ref{eq:IndividSol2} can be evaluated
to assess the underlying statistics. 

It is possible to derive expressions for the parameter variance from
Eqs.~\ref{eq:IndividSol1} to \ref{eq:IndividSol3}: 

\begin{equation}
\left\langle \delta A^2\right\rangle = {\displaystyle 
\left[{\displaystyle \sum\frac{T_{k}^{\,2}}{\sigma_{k}^{2}}}\right]^{-1}}\,;
\label{eq:IndividVar1}
\end{equation}

\begin{equation}
\left\langle \delta B^2\right\rangle =\left[{\displaystyle \sum\frac{1}
{\sigma_{k}^{2}}}\right]^{-1}\,;\label{eq:IndividVar2}
\end{equation}

\begin{equation}
\left\langle \delta C^2\right\rangle =\frac{1}{A^{2}}
\left[\sum\frac{T_{k}'^{2}}{\sigma_{k}^{2}}\right]^{-1}\,.\label{eq:IndividVar3}
\end{equation}

The solution is unbiased, as can be shown by evaluating the expectation
value of Eqs.~\ref{eq:IndividSol1} to \ref{eq:IndividSol3}. The
right-hand terms vanish, so that the expected discrepancy between
true values and their (final) estimates is also zero: 
\begin{equation}
\left\langle \delta A\right\rangle =\left\langle \delta B\right\rangle = 
\left\langle \delta C\right\rangle =0\,.\label{eq:IndividUnbiased}
\end{equation}
The actual discrepancy after a finite number of iterations is investigated
by simulation in the next sections. 

The correlation between parameters can be evaluated in terms of Pearson's
(linear) correlation coefficient formula: 

\begin{equation}
\rho_{A,B}=\frac{\left\langle \delta A\delta B\right\rangle }
{\sqrt{\left\langle \delta A^{2}\right\rangle \left\langle \delta 
B^{2}\right\rangle }} 
= \frac{{\textstyle \sum\nicefrac{T_{k}}{\sigma_{k}^{2}}}}
{\sqrt{{\textstyle \sum\nicefrac{T_{k}^{\,2}}{\sigma_{k}^{2}}}
{\textstyle \sum\nicefrac{1}{\sigma_{k}^{2}}}}}\,;\label{eq:LinCorrCoeff1}
\end{equation}

\begin{equation}
\rho_{A,C}=\frac{\left\langle \delta A\delta C\right\rangle }
{\sqrt{\left\langle \delta A^{2}\right\rangle \left\langle \delta 
C^{2}\right\rangle}} 
= \frac{-{\textstyle \sum\frac{T_{k}T_{k}'}{\sigma_{k}^{2}}}}
{\sqrt{{\textstyle \sum\frac{T_{k}^{\,2}}{\sigma_{k}^{2}}}
{\textstyle \sum\frac{T_{k}'^{2}}{\sigma_{k}^{2}}}}}\,;\label{eq:LinCorrCoeff2}
\end{equation}

\begin{equation}
\rho_{B,C}=\frac{\left\langle \delta B\delta C\right\rangle }
{\sqrt{\left\langle \delta B^{2}\right\rangle \left\langle \delta 
C^{2}\right\rangle }} 
= \frac{-{\textstyle \sum\frac{T_{k}'}{\sigma_{k}^{2}}}}
{\sqrt{{\textstyle \sum\frac{1}{\sigma_{k}^{2}}}
{\textstyle \sum\frac{T_{k}'^{2}}{\sigma_{k}^{2}}}}}\,.\label{eq:LinCorrCoeff3}
\end{equation}

The correlation values shall be investigated in the simulations below. 

A simple qualitative consideration can be derived from the above 
Eqs.~\ref{eq:LinCorrCoeff1}
to \ref{eq:LinCorrCoeff3}: in case of a symmetric signal distribution,
with anti-symmetric derivative (i.e. respectively an even and odd
function), and in circumstances leading to a symmetric readout region
and signal variance, the correlation between photo-centre and either
amplitude and background vanishes. This is due to the summation of
odd functions (a combination of the template, its derivative, and variance)
over a symmetric region, and can be expected to hold approximately
also in case of moderate deviation from the ideal case. In practice,
exact cancellation would not be achieved because residual errors on
the parameters imply computation of the above functions in slightly
off-centered positions, thus departing from symmetry. 
\\ 
The obvious correlation between amplitude and background is retained.

\subsection{Combined Estimate (CE) }
Eqs.~\ref{eq:derivChiSq} are expanded simultaneously for 
all parameters, truncating to the first order under normal 
conditions of smoothness. 

The overall solution must satisfy the system of equations 
\begin{equation}
{\textstyle \delta A\sum\frac{T_{k}^{\,2}}{\sigma_{k}^{2}}}  
{\textstyle +\delta B\sum\frac{T_{k}}{\sigma_{k}^{2}}} 
-A_{E}\,\delta C {\textstyle \sum\frac{T_{k}'T_{k}}{\sigma_{k}^{2}}} 
 = {\textstyle \sum\frac{d_{k}T_{k}}{\sigma_{k}^{2}}}\label{eq:CE_1}
\end{equation}
\begin{equation}
{\textstyle \delta A\sum\frac{T_{k}}{\sigma_{k}^{2}}} 
+\delta B\sum\frac{1}{\sigma_{k}^{2}} 
{\textstyle -A_{E}\,\delta C\sum\frac{T_{k}'}{\sigma_{k}^{2}}} 
 = {\textstyle \sum\frac{d_{k}}{\sigma_{k}^{2}}}\label{eq:CE_2}
\end{equation}
\begin{equation}
{\textstyle \delta A\sum\frac{T_{k}T_{k}'}{\sigma_{k}^{2}}} 
+\delta B\sum\frac{T_{k}'}{\sigma_{k}^{2}} 
{\textstyle {\textstyle -A_{E}\,\delta C\sum\frac{T_{k}'^{2}}{\sigma_{k}^{2}}}} 
 = {\textstyle \sum\frac{d_{k}T_{k}'}{\sigma_{k}^{2}}}\label{eq:CE_3}
\end{equation}
and the explicit solution is defined e.g. by Cramer's rule (still
acceptably efficient with three equations). We may build the system
determinant 

\begin{equation}
\Delta={\textstyle -A_{E}{\textstyle \left|\begin{array}{ccc}
{\textstyle \sum\frac{T_{k}^{2}}{\sigma_{k}^{2}}} & 
{\textstyle \sum\frac{T_{k}}{\sigma_{k}^{2}}} & 
{\textstyle \sum\frac{T_{k}'T_{k}}{\sigma_{k}^{2}}}\\
{\textstyle \sum\frac{T_{k}}{\sigma_{k}^{2}}} & 
{\textstyle {\textstyle \sum\frac{1}{\sigma_{k}^{2}}}} & 
{\textstyle \sum\frac{T_{k}'}{\sigma_{k}^{2}}}\\
{\textstyle \sum\frac{T_{k}T_{k}'}{\sigma_{k}^{2}}} & 
{\textstyle \sum\frac{T_{k}'}{\sigma_{k}^{2}}} & 
{\textstyle {\textstyle \sum\frac{T_{k}'^{2}}{\sigma_{k}^{2}}}}
\end{array}\right|}}\,,\label{eq:CollectDeterm}
\end{equation}
and, for convenience, the auxiliary determinants 
\[
\Delta_{A}=\left|\begin{array}{ccc}
{\textstyle \sum\frac{d_{k}T_{k}}{\sigma_{k}^{2}}} & 
{\textstyle \sum\frac{T_{k}}{\sigma_{k}^{2}}} & 
{\textstyle \sum\frac{T_{k}'T_{k}}{\sigma_{k}^{2}}}\\
{\textstyle \sum\frac{d_{k}}{\sigma_{k}^{2}}} & 
\sum\frac{1}{\sigma_{k}^{2}} & 
{\textstyle {\textstyle \sum\frac{T_{k}'}{\sigma_{k}^{2}}}}\\
{\textstyle \sum\frac{d_{k}T_{k}'}{\sigma_{k}^{2}}} & 
{\textstyle \sum\frac{T_{k}'}{\sigma_{k}^{2}}} & 
{\textstyle \sum\frac{T_{k}'^{2}}{\sigma_{k}^{2}}}
\end{array}\right|
\]
\[
\Delta_{B}=\left|\begin{array}{ccc}
{\textstyle \sum\frac{T_{k}^{2}}{\sigma_{k}^{2}}} & 
{\textstyle \sum\frac{d_{k}T_{k}}{\sigma_{k}^{2}}} & 
{\textstyle \sum\frac{T_{k}'T_{k}}{\sigma_{k}^{2}}}\\
{\textstyle \sum\frac{T_{k}}{\sigma_{k}^{2}}} & 
{\textstyle \sum\frac{d_{k}}{\sigma_{k}^{2}}} & 
{\textstyle \sum\frac{T_{k}'}{\sigma_{k}^{2}}}\\
{\textstyle \sum\frac{T_{k}T_{k}'}{\sigma_{k}^{2}}} & 
{\textstyle \sum\frac{d_{k}T_{k}'}{\sigma_{k}^{2}}} & 
{\textstyle \sum\frac{T_{k}'^{2}}{\sigma_{k}^{2}}}
\end{array}\right|
\]
\[
\Delta_{C}=\left|\begin{array}{ccc}
{\textstyle \sum\frac{T_{k}^{2}}{\sigma_{k}^{2}}} & 
{\textstyle \sum\frac{T_{k}}{\sigma_{k}^{2}}} & 
{\textstyle \sum\frac{d_{k}T_{k}}{\sigma_{k}^{2}}}\\
{\textstyle \sum\frac{T_{k}}{\sigma_{k}^{2}}} & 
{\textstyle \sum\frac{1}{\sigma_{k}^{2}}} & 
{\textstyle \sum\frac{d_{k}}{\sigma_{k}^{2}}}\\
{\textstyle \sum\frac{T_{k}T_{k}'}{\sigma_{k}^{2}}} & 
{\textstyle \sum\frac{T_{k}'}{\sigma_{k}^{2}}} & 
{\textstyle \sum\frac{d_{k}T_{k}'}{\sigma_{k}^{2}}}
\end{array}\right|
\]
so that the variable corrections are 
\begin{eqnarray}
\delta A & = & -A_{E}\frac{\Delta_{A}}{\Delta}\,,\nonumber \\
\delta B & = & -A_{E}\frac{\Delta_{B}}{\Delta}\,,\label{eq:CollectSol}\\
\delta C & = & \frac{\Delta_{C}}{\Delta}\,.\nonumber 
\end{eqnarray}

As above, the solution is unbiased: 
\begin{equation}
\left\langle \delta A\right\rangle =\left\langle \delta B\right\rangle = 
\left\langle \delta C\right\rangle =0\,.\label{eq:CollectUnbiased}
\end{equation}

The computation of parameter variance and covariance is somewhat more
cumbersome (although not exceedingly difficult) than in the previous IE 
case, and its main steps are reported in Appendix~\ref{Appendix1} for the 
interested reader. 

The results of combined and independent solutions are compared in
the simulations detailed in Sec.~\ref{sec:Simulation}.

\subsection{Initial Guess and Stopping Criterion}
\label{sub:IniGuess}A very simple approach is adopted, not taking
into account information from external sources on the currently expected
value of signal amplitude, sky background and/or centre location. 

The background is estimated to be simply a fraction of the minimum
pixel signal level: 
\begin{equation}
B_{E}=\eta\cdot\min\left(S_{k}\right)\,,\,0\le\eta\le1\,.\label{eq:BEstIni}
\end{equation}

The total signal amplitude is then estimated as the integral of the
signal subtracted by such background: 
\begin{equation}
A_{E}=\sum\left(S_{k}-B_{E}\right)\,.\label{eq:AEstIni}
\end{equation}

The photo-centre location is finally estimated as the barycentre,
or first moment, of the signal distribution: 
\begin{equation}
C_{E}=\frac{1}{A_{E}}\sum x_{k}\cdot\left(S_{k}-B_{E}\right)\,.\label{eq:CEstIni}
\end{equation}

As the iterative method works by reducing the square discrepancy $D$
between signal and template in Eq.~\ref{eq:ChiSquare}, a convenient
stopping criterion appears to be the variation $\delta D$ of such
quantity between subsequent iterations. When the amplitude of discrepancy
variation becomes smaller than a suitable acceptance threshold, convergence
is assumed to have been reached. 

The issue of initial guess of parameters, and of a convenient value
for the stopping criterion, is evaluated throughout the simulations.

\section{Simulation Results}
\label{sec:Simulation}
The estimation described above depends on the
signal profile, so that the evaluation of general expressions can
only be done on specific cases. 
To provide an assessment of the performance that may be expected 
in realistic cases, we investigate the results provided by the above 
derivation on a set of data spanning over a significant range of 
instrument response variation (represented by means of optical 
aberrations) and source spectral types (blackbody temperatures). 

\subsection{Simulation Implementation }
The data set was used in previous studies \citep{GaiPASP2013Fit}, 
and is described therein. 
We briefly recall that each effective detected 1D signal, 
labeled LSF for similarity with the corresponding optical Line Spread 
Function, is derived through numeric computation of the diffraction 
integral, providing the Point Spread Function (PSF). 
The PSF is then composed with simple source spectra (blackbodies
at given temperatures) and detection effects (nominal pixel size,
Modulation Transfer Function [MTF] and TDI operation; across scan 
binning). 
The process is replicated for a set of $N_I=10,000$ independent 
instances. 

For simplicity, the data are generated and processed over 12 pixels 
over the whole simulation range, even if this does not match in full 
detail the actual {\it Gaia} operations recalled in Sec.~\ref{sec:GaiaMission}. 
The simulation is run subsequently on the magnitude range 
$11 \, mag \le G \le 21 \, mag$, in steps of $0.25\, mag$, for 
the set of background level values $B = 0, 10, 30, 100$ 
photons/pixel/exposure. 

For each combination of magnitude and background, every LSF instance is 
properly scaled and used as input to the mathematical framework from 
Sec.~\ref{sec:MaxLikApproach}, 
deriving the expected noise level and correlation on amplitude, background 
and location estimate in Sec.~\ref{sec:prediction}. 
Then, the signals are superposed to the appropriate noise level, including 
shot noise from amplitude and background, and a readout noise set to $5$ 
equivalent photons. 
The noisy data are fed to our implementation of the IE and CE algorithms, 
providing experimental results discussed in Sec.~\ref{sec:simulation}. 
We also evaluate the computational implications of our implementation.

\subsection{Predicted Performance }
\label{sec:prediction}
As a first step, we evaluate the expected error level on independent and 
combined solutions, and we derive their correlation coefficients. 
The computation is noiseless in the sense that each instance is defined 
by the current nominal values of parameters, signal template, and variance. 

\begin{figure}
\begin{centering}
\includegraphics[width=0.99\columnwidth,height=0.47\textheight]{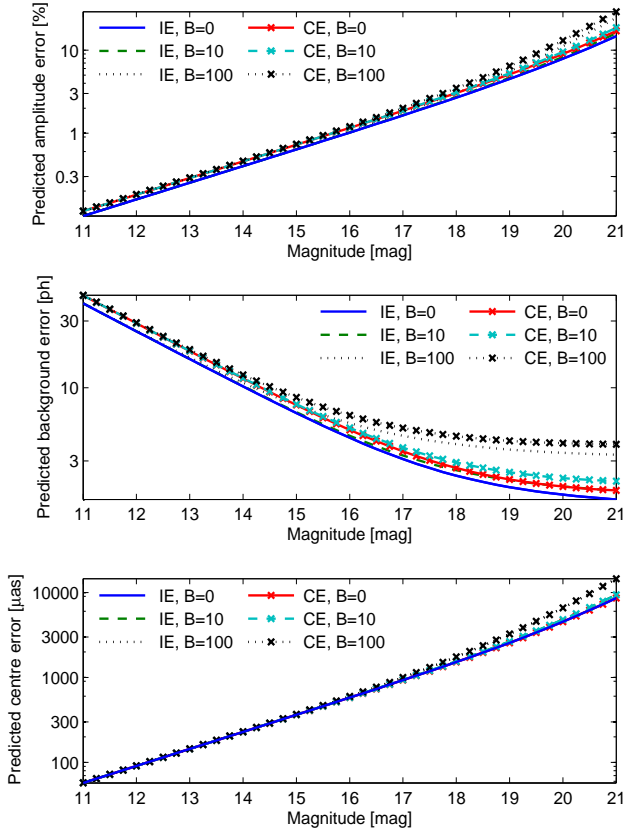}
\par\end{centering}
\caption{\label{fig:EstError1}Noise level, predicted by independent and combined
algorithms, on signal amplitude (top panel), background (middle) and location
(bottom panel)}
\end{figure}

\subsubsection{Predicted Estimate Noise }
We first compare the predictions of the two algorithms in terms of
noise and correlation. 
Intensity values are in photons; location is in
micro-arcsec ($\mu as$). 

In Fig.~\ref{fig:EstError1}, the estimated noise on signal amplitude
$A$ (top panel), background level $B$ (middle panel) and photo-centre
$C$ (bottom panel) is shown, from both independent and combined algorithm,
over the magnitude range, for background level $B=0,\,10,\,100$~photons 
per pixel). 
The overall trends of noise evidence that the performance of either 
algorithm is quite similar, with slightly lower values from IE. 
This is not surprising, as IE assumes implicitly to use all of the 
signal information for the least square estimation of the current 
parameter only, whereas the combined algorithm, more realistically, 
spreads the available information among all of them. 
The ``greedy'' independent algorithm therefore provides a slight 
underestimate of the error.

\begin{figure}
\begin{centering}
\includegraphics[width=0.99\columnwidth,height=0.47\textheight]{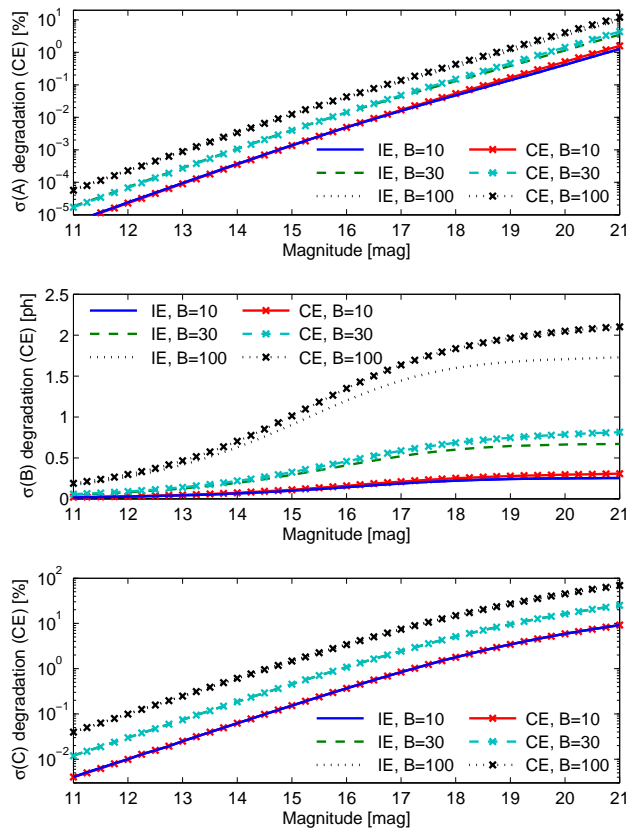}
\par\end{centering}
\caption{\label{fig:DegradBackColl}Relative degradation of parameter 
noise due to background, with respect to the $B=0$ case. }
\end{figure}

The amplitude error (top panel) grows for increasing magnitude, as 
can be expected from natural limitations; moreover, the difference 
between IE and CE noise estimates also slightly increases. 
On the contrary, the noise on background estimate (middle panel) is smaller 
at fainter magnitude, with negligible variation of the difference between 
algorithms. 
This behaviour is understandable, as higher signal levels completely 
``flood'' the background, making it harder to estimate, and inducing 
larger residual noise. 
The location performance from the two algorithms (bottom panel) is basically 
undistinguishable on this scale. 

At the faint end, the error curves evidence larger spread and an increase 
with respect to the near straight line shown at brighter magnitudes, 
dominated by the source photon noise. 
This is not surprising, because at fainter magnitude the overall SNR drops 
and is progressively more affected by higher background levels. 
The effect is small on the scale of Fig.~\ref{fig:EstError1}, therefore 
the variation with respect to the zero background case is shown 
in Fig.~\ref{fig:DegradBackColl},
respectively, for relative amplitude noise 
$[\unit{\sigma\left(A;B\right)/\sigma\left(A;B=0\right)} - 1]$
(top panel), differential background noise 
$[\unit{\sigma\left(B;B\right)-\sigma\left(B;B=0\right)}]$
(middle panel) and relative location error 
$[\unit{\sigma\left(C;B\right)/\sigma\left(C;B=0\right)} - 1]$
(bottom panel). 

\subsubsection{Predicted Correlation Among Parameters }
\begin{figure}
\begin{centering}
\includegraphics[width=0.99\columnwidth,height=0.47\textheight]{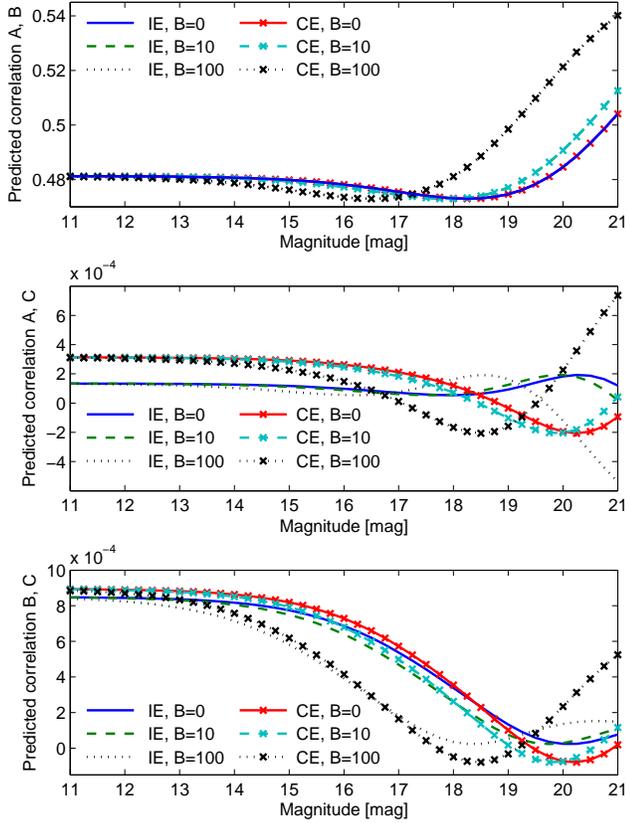}
\par\end{centering}
\caption{\label{fig:EstCorrCoeff1}Predicted correlation between parameters: 
amplitude and background (top); amplitude and location (middle); background
and location (bottom). }
\end{figure}

Over the selected range of magnitude and background, the correlation 
coefficients of errors between parameter pairs are computed, from 
Eqs.~\ref{eq:LinCorrCoeff1} to \ref{eq:LinCorrCoeff3} for IE and from 
the corresponding system of equations for CE; the results are shown 
in Fig.~\ref{fig:EstCorrCoeff1}, 
respectively as the correlation coefficient between amplitude and
background (top panel, sign reversed), amplitude and location (middle 
panel), and background and location (bottom panel). 

The correlation between signal amplitude and background (top panel) 
is not negligible ($\sim 0.5$), and this is not surprising, as 
mentioned in Sec.~2. 
The correlation between signal amplitude and location (middle panel), 
and respectively between location and background (bottom panel), are 
both very low,  and compatible with the sample dispersion (not shown 
in the figure). 
For each term, the difference between algorithms 
is marginal. 
\\ 
The spread among the correlation curves increases at faint magnitudes. 
As for the predicted noise, this can be related to the increasing 
relevance of background level, which acts as a uniform pedestal on 
the signal variance.  
The relative weight of signal and background changes over the simulation 
range, and this affects the results on both noise and correlation. 

\begin{figure}
\begin{centering}
\includegraphics[width=0.99\columnwidth,height=0.47\textheight]{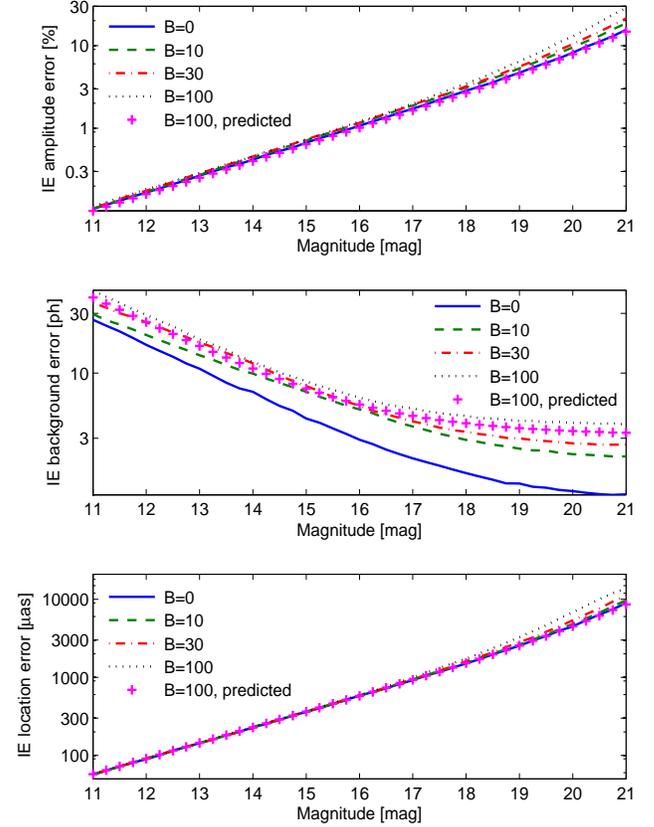}
\par\end{centering}
\caption{\label{fig:SimError1}IE simulation noise on amplitude (top), background
(middle) and location (bottom) with respect to model prediction }
\end{figure}

The performance degradation with increasing background is larger at 
fainter magnitudes (due to decreasing SNR). 
The effects appear to be acceptable 
on the estimation of amplitude (top panel; $1\%$ to $10\%$ increase), 
negligible on background itself (middle panel; $\sim 2$~photons/pixel 
at the faint end), and marginal on location noise (bottom panel) 
down to the faint magnitude range, where the error grows by up to 
a factor two for the worst combination of magnitude and background.

\subsection{Performance on Noisy Data }
\label{sec:simulation}
The initial guess on parameters is generated from the noisy data 
according to the initial guess criterion from Sec.~\ref{sub:IniGuess}. 
The noisy data are then processed by both independent and combined algorithms, 
using the current noiseless LSF as template. 
The stopping criterion corresponds to a limit on square discrepancy variation 
$\delta D \le 1e-3$; the convenience of such choice, based on initial 
experimentation on the algorithm performance, is discussed on the basis of 
the results presented in Sec.~\ref{sub:SquareDiscr}. 

The predicted performance is verified on noisy data using a single noise 
instance for each LSF; the result statistics is therefore a combination of 
noise propagation and sample variability. 
The algorithm performance is discussed in statistical terms on estimated 
parameter noise and discrepancy over the LSF dataset.

\begin{figure}
\begin{centering}
\includegraphics[width=0.99\columnwidth,height=0.47\textheight]{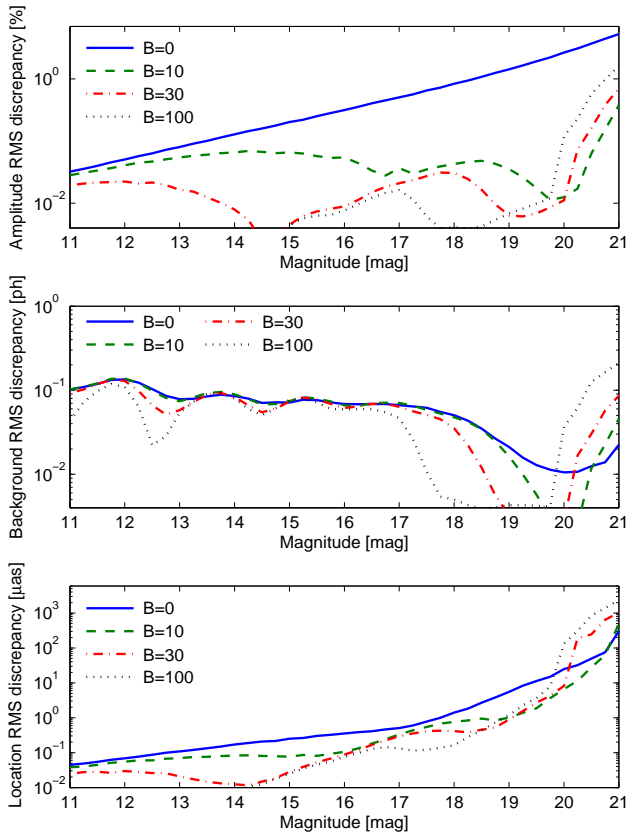}
\par\end{centering}
\caption{\label{fig:SimDiscr1}RMS discrepancy between IE and CE results 
on amplitude (top), background (middle) and location (bottom) }
\end{figure}

\subsubsection{ Estimate Noise from Simulation }
The mean noise on amplitude (top panel), background (middle panel) and 
location (bottom panel),
evaluated as RMS discrepancy with respect to the input (``true'')
values, is shown for IE in Fig.~\ref{fig:SimError1}, with
the corresponding predictions (crosses) for the $B=100$ case 
(Sec.~\ref{sub:IAS}, Eqs.~\ref{eq:IndividVar1} to \ref{eq:IndividVar3}). 

\begin{figure}
\begin{centering}
\includegraphics[width=0.99\columnwidth,height=0.47\textheight]{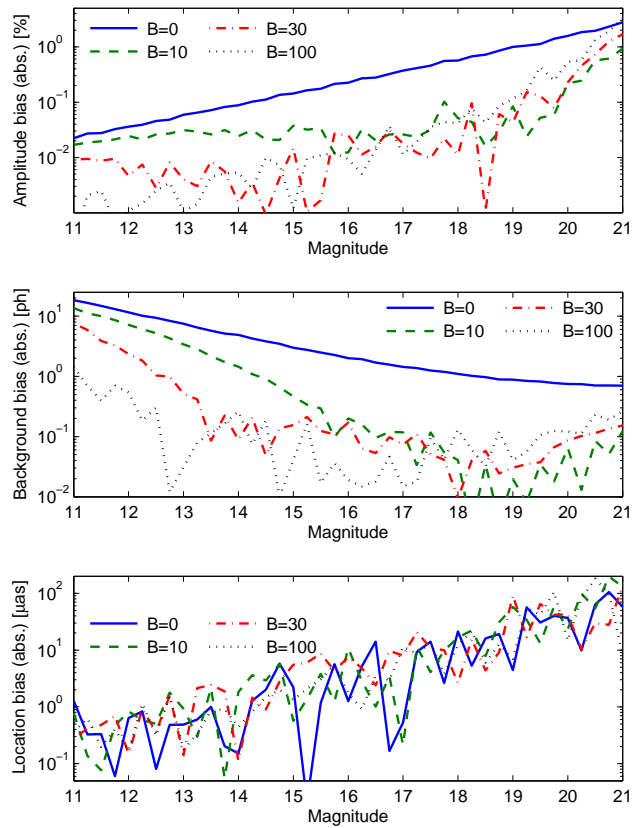}
\par\end{centering}
\caption{\label{fig:SimBias}Mean discrepancy of IE estimates with respect 
to input values on amplitude (top), background (middle) and location (bottom) }
\end{figure}

The discrepancy between predicted and simulated noise increases with 
increasing background and fainter magnitudes, both factors associated 
to decreasing SNR. 
It may be noted that in such conditions the first order expansion used 
in Sec.~\ref{sec:MaxLikApproach} is a progressively less accurate 
approximation.

\subsubsection{Mutual Consistency and Bias }
The numerical results of the two algorithms are compared with each
other; in particular, we focus on the RMS discrepancy between IE and CE 
outputs, shown in Fig.~\ref{fig:SimDiscr1}, 
respectively, for amplitude (top panel), background (middle panel) and 
location (bottom panel).
It appears that the estimates from both algorithms coincide over most of 
the range, within a very small fraction of the noise level of each 
parameter. 
Therefore, CE results were not shown in Fig.~\ref{fig:SimError1}, as 
they would have been mostly indistinguishable from IE values on the plots. 

The mean discrepancy between IE results and the input values, considered
as an indication of systematic estimation error, i.e. bias,
is shown in Fig.~\ref{fig:SimBias}, respectively, for amplitude (top 
panel), background (middle panel) and centre (bottom panel). 
The bias appears to be consistent
with zero, taking into account the noise on each parameter
(Fig.~\ref{fig:SimError1}). 
Given the consistency between IE and CE (Fig.~\ref{fig:SimDiscr1}), 
we can conclude that the latter algorithm provides unbiased estimates 
as well. 

\subsubsection{ Correlation Among Parameters from Simulation } 
\begin{figure}
\begin{centering}
\includegraphics[width=0.99\columnwidth,height=0.47\textheight]{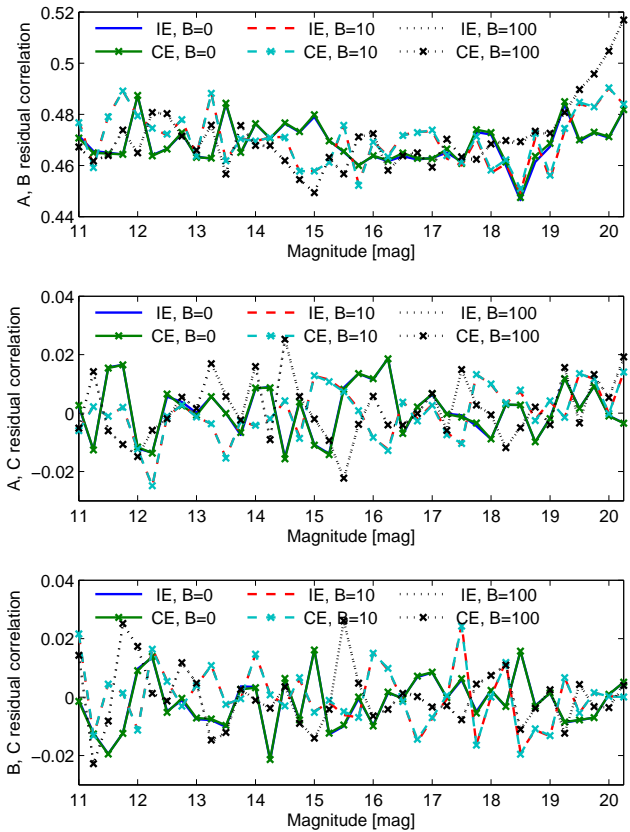}
\par\end{centering}
\caption{\label{fig:SimResids1}Experimental correlation among simulation 
residuals of amplitude vs. background (top), amplitude vs. location (middle), 
and background vs. location (bottom). }
\end{figure}

An experimental assessment of the correlation between estimated parameters 
is also implemented by (a) taking the discrepancy between algorithm 
results and input values for each parameter; (b) evaluating the mean over 
the simulation sample of the product of residuals for each parameter pair; 
and (c) normalising by the corresponding standard deviations. 
\\ 
The residual correlation is shown in Fig.~\ref{fig:SimResids1}, 
retrieving the predicted values, i.e. about $0.5$ 
between flux and background, and close to zero for the other 
combinations. 
Due to the simulation size, the experimental noise on the correlation 
coefficients can be expected to be of order of $1\%$, consistent with 
the fluctuation in the results. 

\subsubsection{Processing Time }
\label{sub:ConvProcTime}
\begin{figure}
\begin{centering}
\includegraphics[width=0.99\columnwidth,height=0.30\textheight]{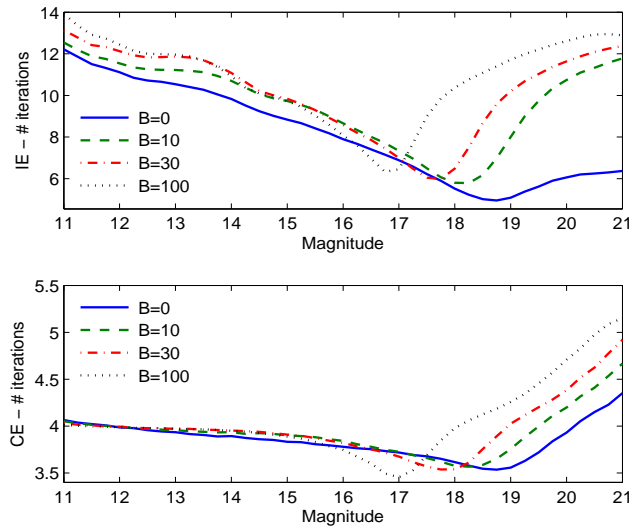}
\par\end{centering}
\caption{\label{fig:SimNIter}Mean number of iterations required for convergence
by IE (top) and CE (bottom) }
\end{figure}

The average number of iterations required
for convergence is shown in Fig.~\ref{fig:SimNIter}, respectively, 
for IE (top) and CE (bottom). 
The IE algorithm requires a larger number of iterations than CE, 
with larger variation over the range of magnitude and background. 
A minimum in the average number of iterations is found for some 
combinations of magnitude and background level. 
\\ 
It may be noted that, with increasing magnitude and decreasing SNR, 
the error on initial guess of the parameters has a natural degradation. 
Besides, at fainter magnitude we expect larger errors on at least 
some parameters (relative amplitude and location), as from 
Fig.~\ref{fig:EstError1}. 
This means that larger residual errors (in absolute terms) are acceptable 
because they induce sufficiently small variations on the 
square discrepancy, thus fitting the stopping criterion. 
The shape of the convergence curves appear to reflect the interplay 
between initial guess and stopping criterion. 

\begin{figure}
\begin{centering}
\includegraphics[width=0.99\columnwidth,height=0.16\textheight]{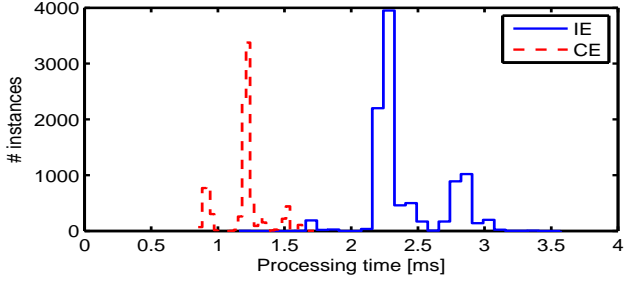}
\par\end{centering}
\caption{\label{fig:ProcTime}Histogram of processing time for IE and CE, 
$G=15\,mag$, $B=10$~photons/pixel}
\end{figure}

The average processing time per LSF instance is larger for IE than
for CE, respectively $2.64\,ms$ against $1.46\,ms$, consistently
with the lower number of iterations required for convergence. 
A histogram of IE and CE processing time per instance, for 
$G = 15\,mag$ and $B = 10$~photons/pixel,
is shown in Fig.~\ref{fig:ProcTime}. 
Given the number of iterations associated to this case 
(Fig.~\ref{fig:SimNIter}), 
roughly twice as large for IE than for CE, it appears that in our case 
the time per iteration is comparable. 

The simulation is implemented in the Matlab environment on a desktop
computer with 3.3~GHz CPU and 8~GB RAM, under 64~bit Windows
operating system. 

\subsubsection{Square Discrepancy and Convergence }
\label{sub:SquareDiscr}
\begin{figure}
\begin{centering}
\includegraphics[width=0.99\columnwidth,height=0.31\textheight]{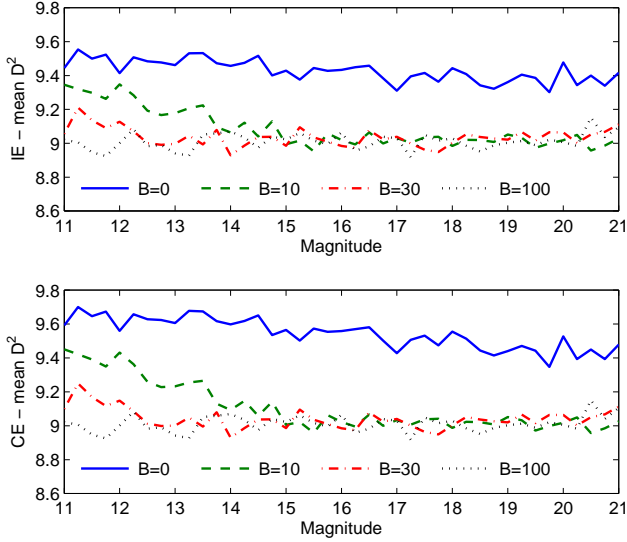}
\par\end{centering}
\caption{\label{fig:SimD2}Average square discrepancy over simulation sample
as a function of magnitude for IE (top) and CE (bottom) }
\end{figure}

The value of square discrepancy at convergence,
averaged over the LSF sample, is shown in Fig.~\ref{fig:SimD2} for
IE (top) and CE (bottom). 
It may be noted that the value is consistent with the expectations on 
$\chi^{2}$, taking into account that the input signal is composed of 12 
samples, and three parameters are estimated. 

\begin{figure}
\begin{centering}
\includegraphics[width=0.99\columnwidth,height=0.58\textheight]{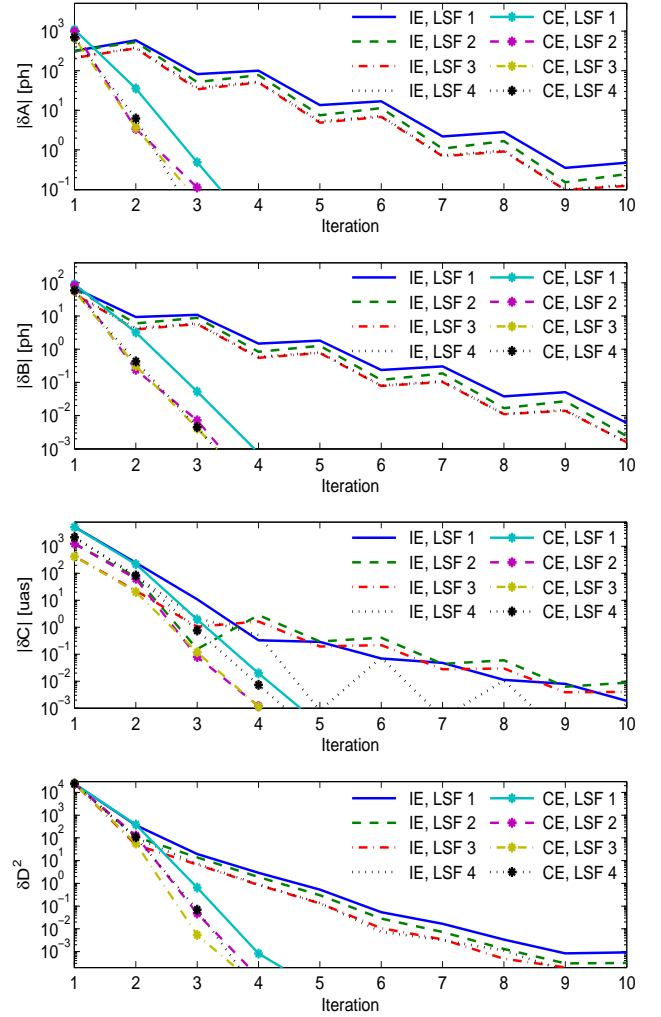}
\par\end{centering}
\caption{\label{fig:CorrIter}Parameter adjustment vs. iterations on four LSF 
instances by IE and CE algorithms. From top to bottom: amplitude, background, 
photo-centre, and square discrepancy. }
\end{figure}

The parameter adjustment trend and the square discrepancy evolution
throughout subsequent iterations of IE and CE algorithms have been 
recorded for a small sample of LSF instances, at a signal amplitude 
corresponding to $G\simeq15\,mag$ and background $B=100$~photons per 
pixel, with centre location close to zero. 
The curves related to the first four LSF instances are shown
in Fig.~\ref{fig:CorrIter}, respectively, for amplitude, background,
photo-centre, and square discrepancy (top to bottom). 

The convergence is much faster for CE than IE; moreover, the CE trend 
is quite monotonic for all parameters. 
Conversely, the IE estimates of $A$ and $B$ corrections evidence 
significant oscillations around a slow decreasing trend, in phase 
opposition. 
We remark that anti-correlation between amplitude and background 
estimates was expected from prime principles, as mentioned in 
Sec.~\ref{sec:MaxLikApproach}. 

The trend of square discrepancy is monotonically decreasing and retains 
an appreciable slope throughout several iterations. 
All curves are in logarithmic units, evidencing a rapid evolution of the 
algorithm. 

The stopping criterion for our main simulation, i.e. $\delta D \le 1e-3$,
has been selected according to the observation that, at this level,
the corrections on all parameters are much smaller that the corresponding
predicted noise level (e.g. shown in Fig.~\ref{fig:EstError1}). 
Therefore, the numerical noise is well
within the intrinsic uncertainty on the measurements. 

An alternative threshold value, as well as different initial 
guesses, are in principle negotiable, depending on implementation
trade-offs.

\section{Discussion: Algorithm Performance }
\label{sec:Discussion}The IE and CE algorithms have been developed under
the concept of an underlying signal model to which measurements are
to be matched in a maximum likelihood approach, thanks to a suitable
choice of parameters: in our case amplitude, background, and location.
The match between model and observations is noise limited, 
provided they are mutually consistent, as ensured by calibration 
and monitoring of the instrument response. 

Hereafter, we discuss some peculiarities of the two algorithms evaluated. 

Both IE and CE appear to provide the same results, with differences (Fig.~\ref{fig:SimDiscr1}) much smaller than the typical error 
(Fig.~\ref{fig:SimError1}) in our simulation. 
Therefore, the choice between algorithms is not suggested 
by considerations on the correctness of their results. 
Our coding was rather straightforward, i.e. not necessarily 
optimised for either IE or CE, and practical performance depends 
on a number of additional implementation aspects: operating system, 
programming language, overall SW infrastructure, availability of 
efficient libraries, and so on. 

In our framework, the CE algorithm (Eqs.~\ref{eq:CE_1} 
to \ref{eq:CE_3}) is more efficient than IE (Eqs.~\ref{eq:IndividSol1} 
to \ref{eq:IndividSol3}), in spite of the 
larger amount of computation required, since it takes less 
iterations, each with comparable processing time, as shown in Sec.~\ref{sub:ConvProcTime}. 
In a different implementation, the relationship between CE and IE 
iteration time may be somewhat different or even reversed, but the 
number of iterations (by deterministic processing of the same 
data through the equations) will be preserved. 

The IE noise prediction (Fig.~\ref{fig:EstError1}) is confirmed
to be somewhat optimistic with respect to the simulation results (Fig.~\ref{fig:SimError1}),
which are more consistent with the CE noise prediction. However, the
discrepancy is small, so that the IE predictions (Eqs.~\ref{eq:IndividVar1} 
to \ref{eq:IndividVar3}) 
can be retained as acceptable ``easy'' estimates. 

\begin{figure}
\begin{centering}
\includegraphics[width=0.99\columnwidth,height=0.24\textheight]{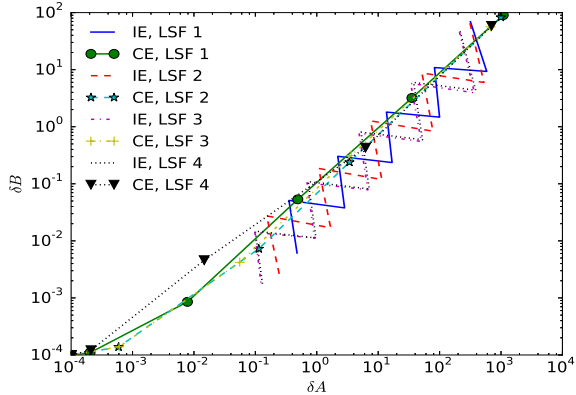}
\par\end{centering}
\caption{\label{fig:Convergence1}A vs. B adjustments throughout 
IE and CE processing for the first four LSF instances (log units) } 
\end{figure}

The rationale of the difference between IE and CE behaviour evidenced 
in Fig.~\ref{fig:CorrIter} can be better understood by a direct comparison 
of $A$ and $B$ adjustments, plotted against each other in 
Fig.~\ref{fig:Convergence1} in logarithmic units. 
The IE parameter adjustments are zigzagging back and forth, whereas the corresponding CE values follow a direct route toward convergence. 

\begin{figure}
\begin{centering}
\includegraphics[width=0.99\columnwidth,height=0.24\textheight]{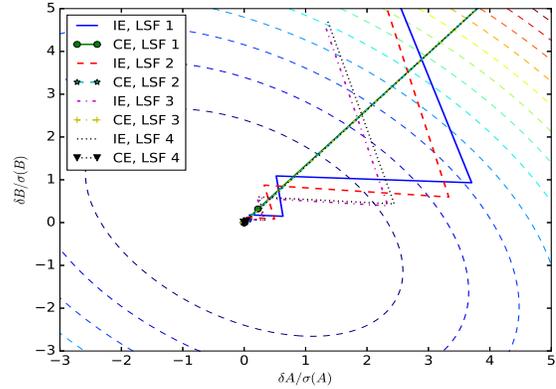}
\par\end{centering}
\caption{\label{fig:Convergence2}A vs. B adjustments superposed to 
square discrepancy contours } 
\end{figure}

It is our understanding that such behavior is related to the correlation 
between flux and background, taken into account by CE and totally ignored 
by IE. 
The correlation effect can be seen more clearly by superposing 
the contour plot of square discrepancy to the $\delta A, \, \delta B$ 
plot, in Fig.~\ref{fig:Convergence2}. 

The flux-background relationship generates a square discrepancy surface 
rotated with respect to the $A, \, B$ axes. 
The CE algorithm operates in a way similar to the gradient descent 
method \citep{NumRec2002}, because 
the shape of the discrepancy function is embedded in the underlying 
mathematical framework. 
Conversely, the IE algorithm acts as a descent method on a single 
coordinate at a time, and cannot therefore aim directly to the 
overall minimum of the error functional. 

The application of the proposed mathematical framework to monitoring 
and diagnostic of relevant signal description parameters related 
to either the astrophysical source or the instrument 
\citep{Busonero2010b,Busonero2012,Busonero2014} 
will be addressed in our future work. 

\section{Conclusion }
We analyse an algorithm for estimation of signal amplitude, background,
and photo-centre location, on 1D data corresponding to
intermediate magnitude {\it Gaia} observations, in the maximum likelihood
framework. The analytical models assess the expected noise level, bias 
and estimate correlation as a function of signal profile, magnitude, 
and background. 

The performance predicted by analytical models has been verified by
simulation over a range of magnitude and background values, on a sample 
of $10,000$ instances of instrument response and source temperature. 
The estimates are unbiased, with low correlation between location and 
either amplitude or background, and well-understood correlation 
between amplitude and background. 
Symmetry properties of the signal further alleviate the 
correlation among results, in particular making the location estimate 
independent of both flux and background. 

The algorithm iterative implementation proves to be effective; 
the suggested choice of parameter initial guess
and stopping criterion appears to work satisfactorily. 
The two algorithm versions tested against each other, respectively,
for independent and combined estimate of the parameters, are
mutually consistent. 
However, CE is significantly more efficient than IE, as convergence 
in the latter case appears to be slowed down by the natural correlation 
between flux and background estimate. 
\vfill

\begin{acknowledgements} 
The activity has been partially funded by the Italian Space Agency (ASI) 
under contracts {\it Gaia} Mission - The Italian Participation to DPAC, 
2014-025-R.1.2015. 
We are grateful to our referee, whose remarks greatly helped in clarifying 
the text and evidencing our results. 
\end{acknowledgements}


\section*{Appendix - Covariance Matrix of the Combined Estimate }
\label{Appendix1} 
The elements of the covariance matrix can be derived from Eqs.~\ref{eq:CE_1}
to \ref{eq:CE_3}, with some manipulations, e.g. squaring each of
them and multiplying them by each other, then taking the expectation
values and using the basic statistics from Eq.~\ref{eq:SigStats}
and Eq.~\ref{eq:CollectUnbiased}. 
We get thus a linear system (Eqs.~\ref{eq:CovMat1} to \ref{eq:CovMat6}) 
in the unknowns $\left\langle \delta A^{2}\right\rangle $,
$\left\langle \delta B^{2}\right\rangle $, $\left\langle \delta C^{2}\right\rangle $,
$\left\langle \delta A\delta B\right\rangle $, $\left\langle \delta A\delta C\right\rangle $ 
and 
$\left\langle \delta B\delta C\right\rangle $. 
\\ 

Eq.~\ref{eq:CE_1} squared: 

\begin{multline}
\left\langle \delta A^{2}\right\rangle \,
\left[\sum\frac{T_{k}^{\,2}}{\sigma_{k}^{2}}\right]^{2}+
\left\langle \delta B^{2}\right\rangle \,
\left[\sum\frac{T_{k}}{\sigma_{k}^{2}}\right]^{2}+\\
A_{E}^{2}\left\langle \delta C^{2}\right\rangle \,
\left[\sum\frac{T_{k}'T_{k}}{\sigma_{k}^{2}}\right]^{2}+
2\left\langle \delta A\delta B\right\rangle \,
\sum\frac{T_{k}^{\,2}}{\sigma_{k}^{2}}\sum\frac{T_{k}}{\sigma_{k}^{2}}+\\
-2A_{E}\left\langle \delta A\delta C\right\rangle \,
\sum\frac{T_{k}^{\,2}}{\sigma_{k}^{2}}\sum\frac{T_{k}'T_{k}}{\sigma_{k}^{2}}+\\
-2A_{E}\left\langle \delta B\delta C\right\rangle \,
\sum\frac{T_{k}}{\sigma_{k}^{2}}\sum\frac{T_{k}'T_{k}}{\sigma_{k}^{2}}=
\sum\frac{T_{k}^{\,2}}{\sigma_{k}^{2}}\,;\label{eq:CovMat1}
\end{multline}

Eq.~\ref{eq:CE_2} squared: 

\begin{multline}
\left\langle \delta A^{2}\right\rangle \,\left[\sum\frac{T_{k}}{\sigma_{k}^{2}}\right]^{2}
+\left\langle \delta B^{2}\right\rangle \,\left[\sum\frac{1}{\sigma_{k}^{2}}\right]^{2}+\\
A_{E}^{2}\left\langle \delta C^{2}\right\rangle \,\left[\sum\frac{T_{k}'}
{\sigma_{k}^{2}}\right]^{2}+
2\left\langle \delta A\delta B\right\rangle \,\sum\frac{T_{k}}{\sigma_{k}^{2}}
\sum\frac{1}{\sigma_{k}^{2}}+\\
-2A_{E}\left\langle \delta A\delta C\right\rangle \,\sum\frac{T_{k}}{\sigma_{k}^{2}}
\sum\frac{T_{k}'}{\sigma_{k}^{2}}+\\
-2A_{E}\left\langle \delta B\delta C\right\rangle \,\sum\frac{1}{\sigma_{k}^{2}}
\sum\frac{T_{k}'}{\sigma_{k}^{2}}=
\sum\frac{1}{\sigma_{k}^{2}}\,;\label{eq:CovMat2}
\end{multline}

Eq.~\ref{eq:CE_3} squared: 

\begin{multline}
\left\langle \delta A^{2}\right\rangle \,
\left[\sum\frac{T_{k}T_{k}'}{\sigma_{k}^{2}}\right]^{2}+\left\langle \delta B^{2}\right\rangle 
\,\left[\sum\frac{T_{k}'}{\sigma_{k}^{2}}\right]^{2}+\\
A_{E}^{2}\left\langle \delta C^{2}\right\rangle \,
\left[\sum\frac{T_{k}'^{2}}{\sigma_{k}^{2}}\right]^{2}+
2\left\langle \delta A\delta B\right\rangle \,
\sum\frac{T_{k}T_{k}'}{\sigma_{k}^{2}}\sum\frac{T_{k}'}{\sigma_{k}^{2}}+\\
-2A_{E}\left\langle \delta A\delta C\right\rangle \,
\sum\frac{T_{k}T_{k}'}{\sigma_{k}^{2}}\sum\frac{T_{k}'^{2}}{\sigma_{k}^{2}}+\\
-2A_{E}\left\langle \delta B\delta C\right\rangle \,
\sum\frac{T_{k}'}{\sigma_{k}^{2}}\sum\frac{T_{k}'^{2}}{\sigma_{k}^{2}}=
\sum\frac{T_{k}'^{2}}{\sigma_{k}^{2}}\,;\label{eq:CovMat3}
\end{multline}

\vfill 

Eq.~\ref{eq:CE_1} $\times$ Eq.~\ref{eq:CE_2} : 

\vfill 


\begin{multline}
\left\langle \delta A^{2}\right\rangle \,
\sum\frac{T_{k}^{\,2}}{\sigma_{k}^{2}}
\sum\frac{T_{k}}{\sigma_{k}^{2}}+\left\langle \delta B^{2}\right\rangle 
\,\sum\frac{T_{k}}{\sigma_{k}^{2}}\sum\frac{1}{\sigma_{k}^{2}}+\\
A_{E}^{2}\left\langle \delta C^{2}\right\rangle \,\sum\frac{T_{k}'T_{k}}{\sigma_{k}^{2}}
\sum\frac{T_{k}'}{\sigma_{k}^{2}}+\\
\left\langle \delta A\delta B\right\rangle \,
{\textstyle 
\left[\sum\frac{1}{\sigma_{k}^{2}}\sum\frac{T_{k}^{\,2}}{\sigma_{k}^{2}}+
\left(\sum\frac{T_{k}}{\sigma_{k}^{2}}\right)^{2}\right]+}\\
-A_{E}\left\langle \delta A\delta C\right\rangle \,
{\textstyle 
\left[\sum\frac{T_{k}^{\,2}}{\sigma_{k}^{2}}\sum\frac{T_{k}'}{\sigma_{k}^{2}}+
\sum\frac{T_{k}}{\sigma_{k}^{2}}\sum\frac{T_{k}'T_{k}}{\sigma_{k}^{2}}\right]}+\\
-A_{E}\left\langle \delta B\delta C\right\rangle \,
{\textstyle 
\left[\sum\frac{T_{k}}{\sigma_{k}^{2}}\sum\frac{T_{k}'}{\sigma_{k}^{2}}+
\sum\frac{1}{\sigma_{k}^{2}}\sum\frac{T_{k}'T_{k}}{\sigma_{k}^{2}}\right]}=\\
\sum\frac{T_{k}}{\sigma_{k}^{2}}\,;\label{eq:CovMat4}
\end{multline}

Eq.~\ref{eq:CE_1} $\times$ Eq.~\ref{eq:CE_3} : 

\begin{multline}
\left\langle \delta A^{2}\right\rangle \,\sum\frac{T_{k}^{\,2}}{\sigma_{k}^{2}}
\sum\frac{T_{k}T_{k}'}{\sigma_{k}^{2}}+\left\langle \delta B^{2}\right\rangle \,
\sum\frac{T_{k}}{\sigma_{k}^{2}}\sum\frac{T_{k}'}{\sigma_{k}^{2}}+\\
A_{E}^{2}\left\langle \delta C^{2}\right\rangle \,\sum\frac{T_{k}'T_{k}}{\sigma_{k}^{2}}
\sum\frac{T_{k}'^{2}}{\sigma_{k}^{2}}+\\
\left\langle \delta A\delta B\right\rangle \,
{\textstyle 
\left[\sum\frac{T_{k}'}{\sigma_{k}^{2}}
\sum\frac{T_{k}^{\,2}}{\sigma_{k}^{2}}+\sum\frac{T_{k}T_{k}'}{\sigma_{k}^{2}}
\sum\frac{T_{k}}{\sigma_{k}^{2}}\right]+}\\
-A_{E}\left\langle \delta A\delta C\right\rangle \,
{\textstyle 
\left[\left(\sum\frac{T_{k}T_{k}'}{\sigma_{k}^{2}}\right)^{2}+
\sum\frac{T_{k}^{\,2}}{\sigma_{k}^{2}}\sum\frac{T_{k}'^{2}}{\sigma_{k}^{2}}\right]}+\\
-A_{E}\left\langle \delta B\delta C\right\rangle \,
{\textstyle 
\left[\sum\frac{T_{k}}{\sigma_{k}^{2}}\sum\frac{T_{k}'^{2}}{\sigma_{k}^{2}}+
\sum\frac{T_{k}'}{\sigma_{k}^{2}}\sum\frac{T_{k}'T_{k}}{\sigma_{k}^{2}}\right]}=\\
\sum\frac{T_{k}'T_{k}}{\sigma_{k}^{2}}\,;\label{eq:CovMat5}
\end{multline}

Eq.~\ref{eq:CE_2} $\times$ Eq.~\ref{eq:CE_3} : 

\begin{multline}
\left\langle \delta A^{2}\right\rangle \,\sum\frac{T_{k}}{\sigma_{k}^{2}}
\sum\frac{T_{k}T_{k}'}{\sigma_{k}^{2}}+\left\langle \delta B^{2}\right\rangle \,
\sum\frac{1}{\sigma_{k}^{2}}\sum\frac{T_{k}'}{\sigma_{k}^{2}}+\\
A_{E}^{2}\left\langle \delta C^{2}\right\rangle \,\sum\frac{T_{k}'}{\sigma_{k}^{2}}
\sum\frac{T_{k}'^{2}}{\sigma_{k}^{2}}+\\
\left\langle \delta A\delta B\right\rangle \,
{\textstyle 
\left[\sum\frac{T_{k}'}{\sigma_{k}^{2}}
\sum\frac{T_{k}}{\sigma_{k}^{2}}+\sum\frac{T_{k}T_{k}'}{\sigma_{k}^{2}}
\sum\frac{1}{\sigma_{k}^{2}}\right]}+\\
-A_{E}\left\langle \delta A\delta C\right\rangle \,
{\textstyle 
\left[\sum\frac{T_{k}}{\sigma_{k}^{2}}
\sum\frac{T_{k}'^{2}}{\sigma_{k}^{2}}+\sum\frac{T_{k}T_{k}'}{\sigma_{k}^{2}}
\sum\frac{T_{k}'}{\sigma_{k}^{2}}\right]}+\\
-A_{E}\left\langle \delta B\delta C\right\rangle \,
{\textstyle 
\left[\sum\frac{1}{\sigma_{k}^{2}}\sum\frac{T_{k}'^{2}}{\sigma_{k}^{2}}+
\left(\sum\frac{T_{k}'}{\sigma_{k}^{2}}\right)^{2}\right]} = \\
\sum\frac{T_{k}'}{\sigma_{k}^{2}}\,.\label{eq:CovMat6}
\end{multline}

The system is based on known information (signal template $T$ and 
variance $\sigma^{2}$, 
nominal flux estimate $A_{E}$), and it can be solved with standard techniques
to provide the terms of the covariance matrix. 
As the solutions are unbiased (Eq.~\ref{eq:CollectUnbiased}), 
the former three unknowns are the variances of the flux, background, 
and photo-centre estimates. 
The latter three unknowns are related to the Pearson's correlation 
coefficients according to expressions similar to those in 
Eqs.~\ref{eq:LinCorrCoeff1} to \ref{eq:LinCorrCoeff3}.


\vfill 

\bibliographystyle{aasjournal} 

\end{document}